%% file: YearsleyDICE2012.tex
\documentclass[a4paper]{jpconf}
\usepackage{graphicx}

\usepackage{graphicx}
\usepackage{dcolumn}
\usepackage{bm}
\usepackage{verbatim}
\usepackage{epsfig}
\usepackage{epstopdf}
\usepackage{amsmath}

\newcommand\beq{\begin{equation}}
\newcommand\eeq{\end{equation}}
\newcommand\bea{\begin{eqnarray}}
\newcommand\eea{\end{eqnarray}}

\def\s{{\sigma}}

\def\half{\frac {1} {2}}
\def\mathbb{}

\begin{document}
\title{An Introduction to the Quantum Backflow Effect}

\author{JM Yearsley$^{1}$ and JJ Halliwell$^{2}$}

\address{$^{1}$ Centre for Quantum Information and Foundations, DAMTP, Centre for Mathematical Sciences, University of Cambridge, Wilberforce Road, Cambridge CB3 0WA, UK}
\address{$^{2}$ Theoretical Physics Group, Blackett Laboratory, Imperial College, London SW7 2BZ, UK }

\ead{jmy27@cam.ac.uk}

\begin{abstract}
We present an introduction to the backflow effect in quantum mechanics -- the phenomenon
in which a state consisting entirely of positive momenta may have negative current
and the probability flows in the opposite direction to the momentum.
We show that the effect is present even for simple states consisting of superpositions of gaussian wave packets, although the size of the effect is small.
Inspired by the numerical results of Penz et al, 
we present a wave function whose current at any time may be computed analytically and which has periods of significant backflow, with a backwards flux equal to about 70 percent of the maximum possible backflow, a dimensionless number $c_{bm} \approx 0.04 $, discovered by Bracken and Melloy. 
This number has the unusual property of being independent of $\hbar$ (and also of all other parameters of the model), despite corresponding to a quantum-mechanical effect, and we shed some light on this surprising property by considering the classical limit of backflow.
We conclude by discussing a specific measurement model in which backflow may be identified in certain measurable probabilities.
\end{abstract}

\section{Introduction}

A striking but little-appreciated phenomenon in quantum mechanics
is the backflow effect. This is the fact that, for a free particle described by a wave function centred in $x<0$ consisting entirely of positive momenta, the probability of remaining
in $x<0$ may nevertheless {\it increase} with time. That is, the quantum-mechanical current at the origin can be negative and the probability flows ``backwards''.

This effect was first noted by Allcock in his early work on the arrival time problem in quantum theory \cite{All}, but was not studied in detail until 1994 when Bracken and Melloy carried out the first systematic study \cite{BrMe}. Their most important discovery was that there exists a limit on the total amount of backflow. Backflow means that the probability of remaining in $x<0$ may increase with time, but this increase is bounded by an amount $c_{bm}$, a dimensionless number computed numerically to be approximately $0.04$. Furthermore although backflow is clearly a nonclassical effect, the quantity $c_{bm}$ is independent of $\hbar$ (and also of the mass of the particle and the time duration of the effect.) For this reason $c_{bm}$ has been described as a new quantum number. 

The effect has been further investigated by a number of authors.
Better estimates for $c_{bm}$ were given in Refs.\cite{Eveson,Penz}, with the latter also giving a numerical estimate of the maximum backflow state.  Some analytic examples of backflow states and the measurability of the effect were explored in Ref.\cite{Muga}. Connections between backflow and the phenomenon of superoscillations were noted in Ref.\cite{Ber}. The effect has also been explored in the Dirac equation \cite{BrMe2}, for a particle in a linear potential \cite{BrMe3} and for angular momentum \cite{strange}. Most recently the present authors have investigated a number of aspects of the backflow effect, including the classical limit \cite{HaYeBF}.
Finally, the effect is often noted in connection with the arrival time problem in quantum theory
\cite{time}

Apart from the studies cited above, there has been little discussion of the effect in the literature to date. The purpose of this contribution is to provide an introduction to the backfow effect both as a reference for researchers encountering it in their studies of quantum theory and also hopefully as an invitation to encourage others to take an interest in the effect.

We begin in Section 2 with a detailed formulation of the problem. We define the current and the flux and consider the properties of the spectrum of the flux operator, which allows us to give the clearest statement of the backflow effect.

In Section 3 we give a simple example of a state with backflow consisting of a superposition of gaussian states. The backflow for such a state is, however, rather small.

In Section 4, we review the numerical computation of the maximal backflow state and eigenvalue.
We also present an analytic wave function with backflow which appears to match closely the numerical solutions for the maximal backflow state by Penz et al \cite{Penz}.
The current at arbitrary times for this wave function may be computed analytically and we find that it has a backflow of approximately 70 percent of the maximal value. This is a much larger backflow than any analytically tractable states previously discovered. 

In Section 5, we consider the naive classical limit $\hbar \rightarrow 0$ of backflow, and in particular, we address the fact that the bound on backflow $c_{bm}$ discovered by Bracken and Melloy appears to be independent of $\hbar$. We show that the expected dependency on $\hbar$ reappears in realistic measurement models, where measurements are described not by exact projectors but by quasiprojectors involving parameters characterizing the imprecision of real measurements. Under these conditions the naive classical limit is restored.

In Section 6, we consider a simple measurement model and discuss the ways in which backflow may be seen in the probabilities for measurements.


We summarize and conclude in Section 7.

\section{Detailed Formulation of the Problem}

\subsection{The Flux}

We consider a free particle with initial wave function $ \psi (x)$ concentrated in $x<0$
and consisting entirely of positive momenta. We consider the probability flux
$F(t_1,t_2)$ crossing the origin during the time interval $[t_1,t_2]$, defined by
\bea
F(t_1, t_2 ) &=& \int_{-\infty}^0 dx \left| \psi (x,t_1) \right|^2 -  \int_{-\infty}^0 dx \left| \psi (x,t_2) \right|^2\nonumber
\\
&=& \int_{t_1}^{t_2} dt \ J(t)
\label{flux}
\eea
where $J(t)$ is the usual quantum-mechanical current at the origin
\beq
J(t) = - \frac{i \hbar}{2m}\left(\psi^{*}(0,t)\frac{\partial
\psi(0,t)}{\partial x}-\frac{\partial \psi^{*}(0,t)}{\partial
x} \psi(0,t)\right)\label{cur}
\eeq
The flux may also be written in terms of the Wigner function \cite{Wig} at time t, $ W_t(p,q)$,
\beq
F(t_1, t_2) = \int_{t_1}^{t_2} dt \int dp dq \ \frac {p} {m} \delta (q) W_t(p,q)
\label{Wig}
\eeq
(For a useful review of the properties of the current and related phase space
distribution functions, see Ref.\cite{Wig2}).
It is useful to write these expressions in an operator form. We introduce projection operators onto the positive and negative $x$-axis, $P = \theta (\hat x)$, and, $ \bar P = 1 - P = \theta (- \hat x)$ respectively. The flux may then be written in terms of the operator $\hat F (t_1, t_2) $
defined by
\bea
\hat F(t_1,t_2) &=& P(t_2) - P(t_1) =  \int_{t_1}^{t_2} dt \ \dot P(t)
\nonumber \\
&=& \int_{t_1}^{t_2} dt \ \frac {i} {\hbar} [H, \theta ( \hat x) ]= \int_{t_1}^{t_2} dt  \hat J(t)
\label{fluxop}
\eea
where the current operator is given by
\beq
\hat J = \frac {1} {2 m} \left( \hat p \delta (\hat x) + \delta (\hat x) \hat p \right)
\label{curop}
\eeq
So Eq.(\ref{flux}) may also be written
\bea
F(t_1, t_2) &=& \langle \hat F(t_1, t_2) \rangle
 =\int_{t_1}^{t_2} dt \ \langle \psi | \hat J(t) | \psi \rangle.
\label{flux2}
\eea
The flux Eq.(\ref{flux}) is a difference between two probabilities and is positive when those probabilities behave according to classical intuition, i.e., when the probability of remaining
in $x<0$ decreases monotonically.


As indicated, in the full quantum-mechanical case, the flux can be negative. One way to see this is to note that the Wigner function need not be positive for general states \cite{Wig}, and thus Eq.(\ref{Wig}) need not be either. 
Another way is from the current operator Eq.(\ref{curop}): both $\hat p$ and $\delta (\hat x)$ are non-negative operators on states with
positive momentum, but since they do not commute, the current operator $\hat J$ is not a positive operator.

\subsection{Most Negative Flux as an Eigenvalue Problem}

Following Bracken and Melloy \cite{BrMe}, a useful way to investigate the backflow effect is to look at the spectrum of the flux operator Eq.(\ref{fluxop}) (restricted to positive momenta). We thus look for solutions to the eigenvalue problem
\beq
\theta ( \hat p) \hat F(t_1, t_2) | \Phi \rangle = \lambda | \Phi \rangle
\label{evalue}
\eeq
where the states $| \Phi \rangle $ consist only of positive momenta.
(We choose an opposite sign convention to \cite{BrMe} so that the backflow states have $\lambda < 0 $). The most negative value of the flux $F(t_1,t_2)$ is then given by the
most negative eigenvalue of Eq.(\ref{evalue}).

By time evolving the state we may choose the time interval $[t_1,t_2]$ to be $[-T/2,T/2]$ and the eigenvalue equation in momentum space then reads
\beq
\frac {1} {\pi} \int_0^\infty dk \ \frac { \sin [(p^2 - k^2) T/ 4 m \hbar ] } {(p-k)} \
\Phi (k)
 = \lambda \Phi (p) \label{evalue1.1}
\eeq
Defining rescaled variables $u$ and $v$ by $ p = 2 \sqrt{ m \hbar / T} u$
and $ k = 2 \sqrt{ m \hbar / T} v$ Eq.(\ref{evalue1.1}) becomes
\beq
\frac {1} {\pi} \int_0^\infty dv \ \frac { \sin (u^2 - v^2)  } {(u-v)} 
\phi (v)
 = \lambda \phi (u)
\label{evalue2}
\eeq
where $ \phi (u) = (m \hbar / 4 T)^{1/4} \Phi (p)$ and is dimensionless. Crucially all physical constants have dropped out of this equation so that the $\lambda$ are dimensionless
and independent of $\hbar, m$ and $T$.

This eigenvalue equation has been studied by a number of authors, both numerically and analytically \cite{BrMe,Eveson,Penz}. 
The eigenvalues lie in the range
\beq
 - c_{bm} \le \lambda \le 1
\label{range}
\eeq
where $c_{bm}$ was computed numerically and found to be
\beq
c_{bm} \approx 0.038452.
\label{cbm}
\eeq
It was conjectured in Ref.\cite{BrMe} that the spectrum is discrete in the interval $[-c_{bm},0]$ but continuous in the interval $[0,1]$. The extremizing state was given numerically by Penz et al \cite{Penz}. 
We will display this maximising state and also a good analytic expression approximating the numerical results in what follows.
%
%
%
%
At present there is no analytic account of the properties  Eqs.(\ref{range}), (\ref{cbm}).
%
Note that the eigenvalues are independent of $T$ which means that the duration of a period of backflow can be arbitrarily long.



\section{Backflow for superpositions of gaussians}\label{secgauss}


In this section we show that the backflow effect arises in the familiar, and also potentially experimentally realisable, setting of a superposition of two gaussian wavepackets.
%
However gaussian wavepackets have support on both positive and negative momentum, so  we will also have to show that this is not the result of any initially negative momentum. We will see that whilst superpositions of gaussian states do indeed give rise to backflow, the size of the effect is considerably smaller than the theoretical maximum. In this Section we work in units in which $\hbar = 1$
and we set the particle mass $m=1$.

Bracken and Melloy demonstrated that the backflow effect may be observed in the simple case of a superposition of two plane waves \cite{BrMe}. This state is unnormalisatble, but
it can be turned into a more physical state by replacing the plane waves with gaussians tightly peaked in momentum, without affecting the basic conclusion that the state displays backflow for well chosen values of the various parameters. 

Consider the normalised state
\beq
\psi(x,t)=\sum_{k=1,2}A_{k}\frac{1}{\sqrt{4\s^{2}+2it}}\exp\left(i p_{k}(x-p_{k}t)-\frac{(x-p_{k}t)^{2}}{4\s^{2}+2it}\right)\label{jsec3}.
\eeq
This is a sum of two initial gaussian wavepackets with equal spatial width $\s$, evolved for a time $t$. If we let $\s\to\infty$ we essentially recover a sum of two plane waves.  Instead of giving the rather complex expression for the current, we will simply plot the current at the origin and the probability of remaining in $x<0$ as functions of time for the state in Eq.(\ref{jsec3}) and for the following set of parameters;
\beq
p_{1}=0.3, \quad p_{2}=1.4,\quad \s=10,\quad A_{1}=1.8,\quad A_{2}=1.
\label{jsec5}
\eeq
We clearly see from these plots that there are several intervals during which the current is negative. 
A magnification of one of these backflow region is shown in Fig.(\ref{jfig2.2}).
The effect is robust with respect to small changes to these parameters.
%
%
%
\begin{figure}[htbp]
  \begin{minipage}[t]{0.47\linewidth}
    \centering
    \includegraphics[width=\linewidth]{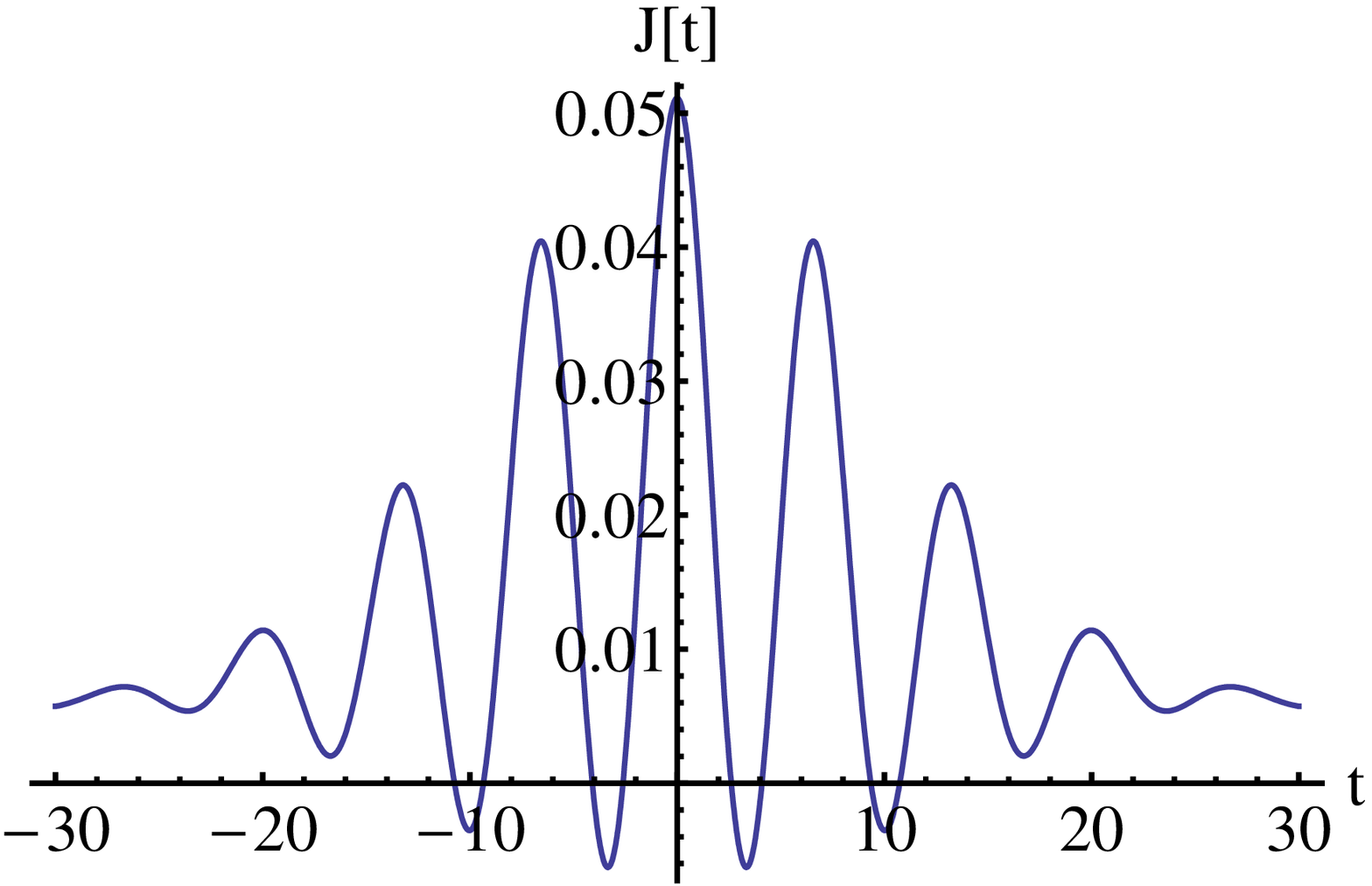}
    \caption{Plot of the current for a wavefunction consisting of a superposition of two gaussians, with the parameters given in Eq.(\ref{jsec5}).}
    \label{jfig2}
  \end{minipage}
  \hspace{.5cm}
  \begin{minipage}[t]{0.47\linewidth}
    \centering
    \includegraphics[width=\linewidth]{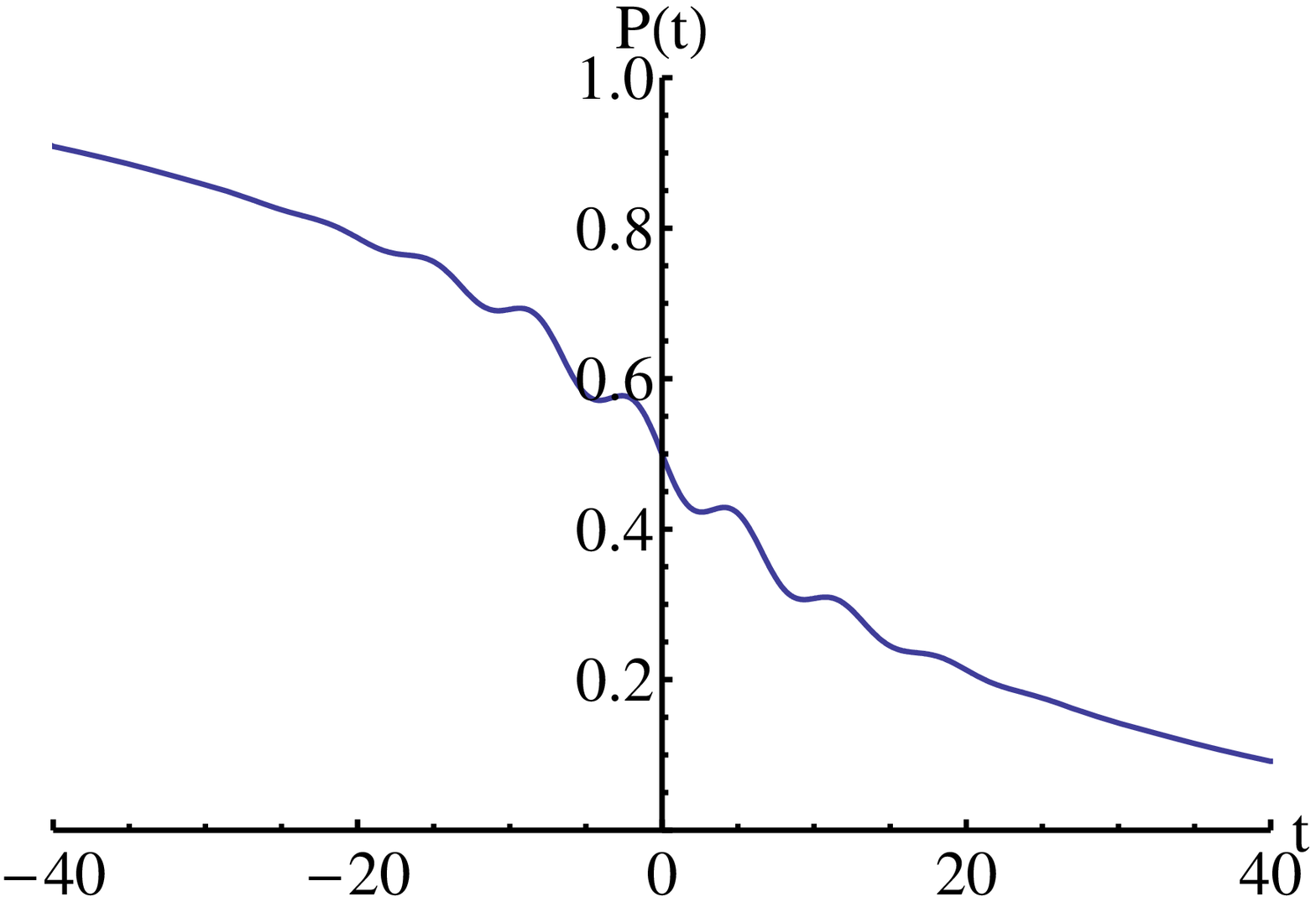}
    \caption{Plot of the probability for remaining in $x<0$ for a wavefunction consisting of a superposition of two gaussians, with the parameters given in Eq.(\ref{jsec5}).}
    \label{jfig2.1}
  \end{minipage}
\end{figure}
\begin{figure}[htbp]
 
  \hspace{0.25\linewidth}
  \begin{minipage}[t]{0.47\linewidth}
    \centering
    \includegraphics[width=\linewidth]{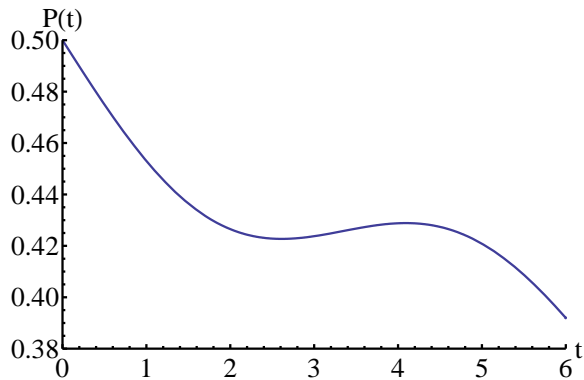}
    \caption{Close up of Fig.(\ref{jfig2.1}). $P(t)$ is clearly seen to increase between $t\approx2$ and $t\approx4$.}
    \label{jfig2.2}
  \end{minipage}
 \hspace{0.25\linewidth}
\end{figure}

The parameters in Eq.(\ref{jsec5}) give rise to the greatest amount of backflow we have been able to find, although we have not  searched the entire parameter space.  The value of the flux during the largest period of backflow is
\beq
F=\int_{t_{1}}^{t_{2}}dt J(t),
\eeq
where $J(t)$ is the current, Eq.(\ref{cur}), and the interval $[t_{1},t_{2}]$ is chosen such that the current is negative for the whole of this time. Computing the flux during this interval gives,
\beq
F\approx -0.0061,\label{jsec6}
\eeq
or about 16 percent of $c_{bm}$.

It is important to check that this probability backflow cannot be explained by the tiny probability of having negative momentum in this gaussian state. It can be shown that the probability that a measurement of the momentum of this state would yield a negative answer is of order $10^{-10}$, so the negative flow of probability is entirely due to the backflow effect \cite{HaYeBF}.

\section{An approximation to the backflow maximising state}

Backflow states may be found by solving the eigenvalue equation Eq.(\ref{evalue2}).
The numerical work of Penz et al \cite{Penz} yielded an approximate eigenstate satisfying Eq.(\ref{evalue2}) giving the most negative eigenvalue $-c_{bm}$, i.e., the largest amount of backflow. 
It is of interest to find analytic expressions for wave functions matching these numerical results closely and for which the current may be computed analytically. We give such a state in this section. To be clear, we are not giving an approximate analytic solution to the eigenvalue problem, Eq.(\ref{evalue2}). Rather, inspired by the numerical solution, we will exhibit an analytic expression for a wave function which closely matches the backflow maximising state, and show that is has significant negative flux.

\subsection{Numerical results}

\begin{figure}[htbp]
  \begin{minipage}[t]{0.47\linewidth}
    \centering
    \includegraphics[width=\linewidth]{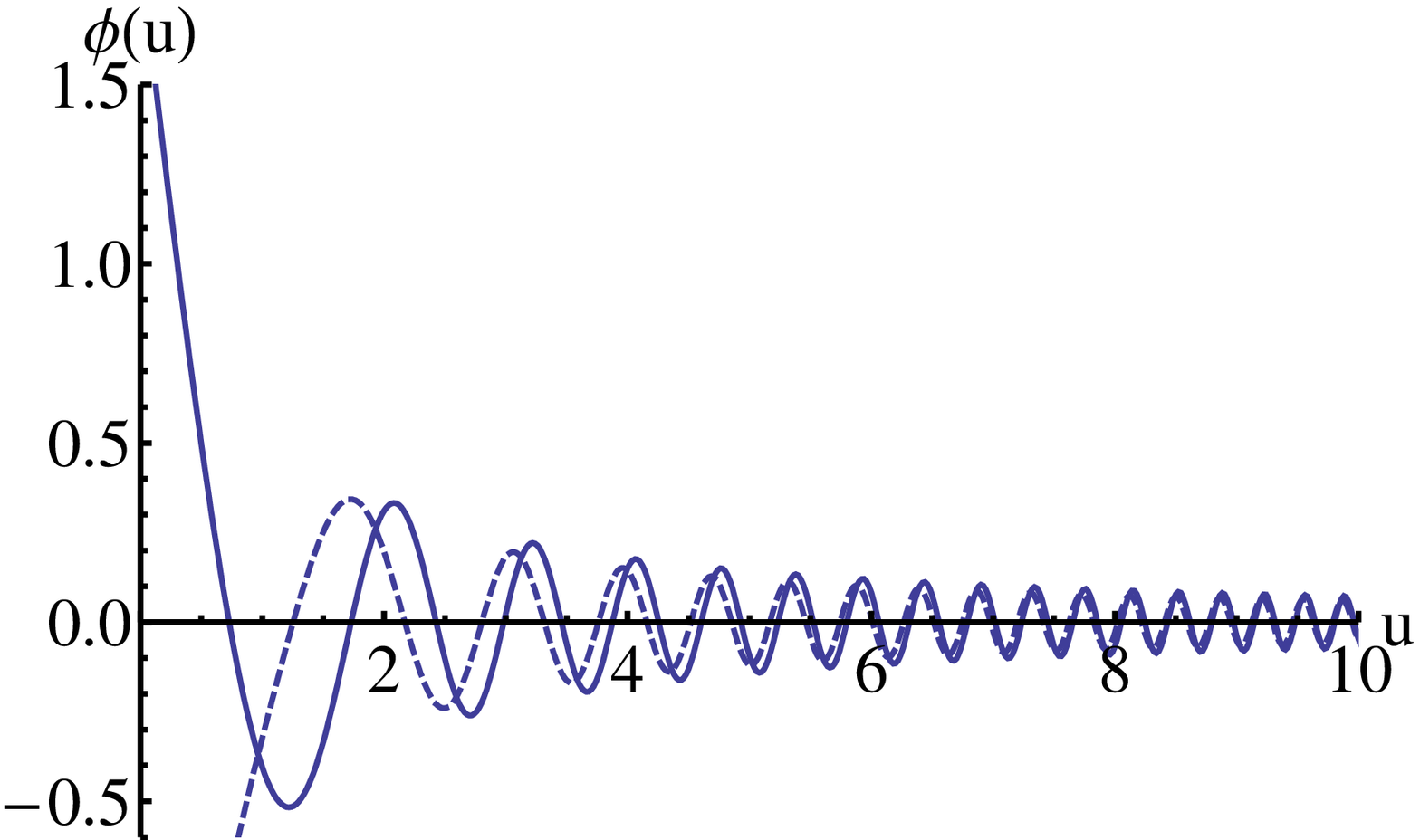}
   \caption{Plot of $\phi_{max}$ (solid line) together with $\phi_{as}$ (dashed line).}
\label{figphiexact}
  \end{minipage}
  \hspace{.5cm}
  \begin{minipage}[t]{0.47\linewidth}
    \centering
    \includegraphics[width=\linewidth]{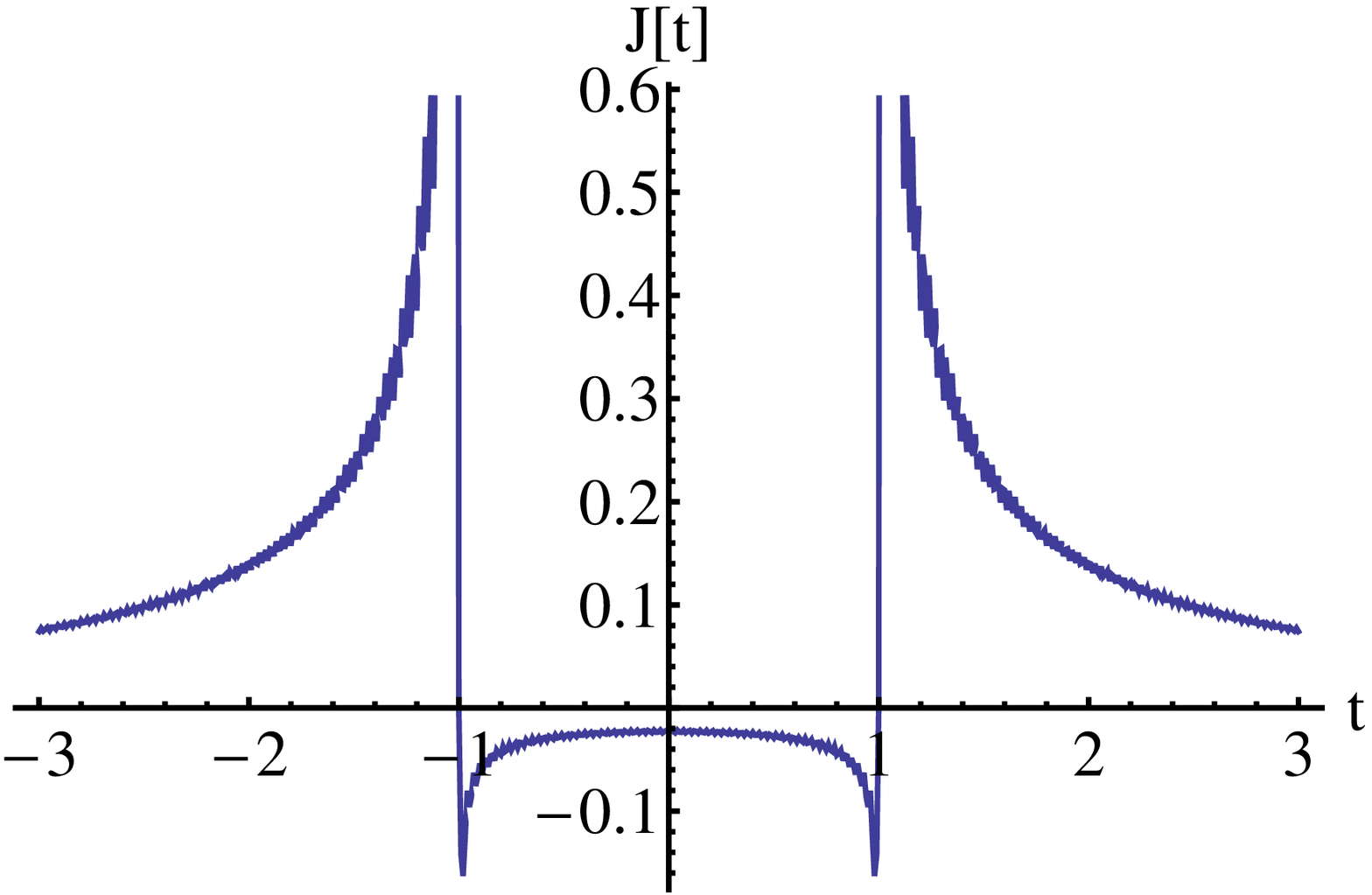}
  \caption{The current, $J(t)$, as computed from $\phi_{max}$.}
\label{figjexact}
  \end{minipage}
\end{figure}

We first review the numerical results of the computation of the backflow state.
We have repeated the numerical analysis of Penz et al.\cite{Penz} of the optimizing state, $\phi_{max}$, and its current in order to compare with our analytic approximation.
In Fig.(\ref{figphiexact}) we plot the numerically computed maximum backflow state, $\phi_{max}$, together with the function
\beq
  \phi_{as}(u)=-\frac{1}{10} \frac{\cos(u^{2})}{u},
  \eeq
which seems to match well the asymptotic form of $\phi_{max}$. 

We plot in Fig.(\ref{figjexact}) the current computed from $\phi_{max}$.
Note that the current appears to have a specific singularity structure at $t= \pm 1$, where it jumps between $\pm \infty$. This is presumably related to some properties of the flux operator, but it also seems to depend on the asymptotic behavior of $\phi_{max}$ \cite{future}. 
These two plots are in agreement, in general shape, with the numerical results of Penz et al \cite{Penz} and we will compare our analytic results with these plots in what follows.

\subsection{Analytic Approximation}

Consider the momentum space wavefunction
\beq
\phi_{A}(u)=N\left[a e^{-b u}+(\frac{1}{2}-C(u))\right],\quad a,b\in\mathbb{R}
\label{guess2}
\eeq
where $N$ is a normalisation factor. Here
\beq
C(u)=\mbox{FresnelC}\left(\sqrt{{\frac{2}{\pi}}}u\right)=\sqrt{\frac{2}{\pi}}\int_{0}^{u}dx\cos(x^{2}).
\eeq
This has the asymptotic form,
\beq
\phi_A(u)\sim N \frac{\sin(u^2)}{u}.
\eeq
Note that this does not in fact match the asymptotic form of $\phi_{max}$.

\begin{figure}[htbp]
  \begin{minipage}[t]{0.47\linewidth}
    \centering
    \includegraphics[width=\linewidth]{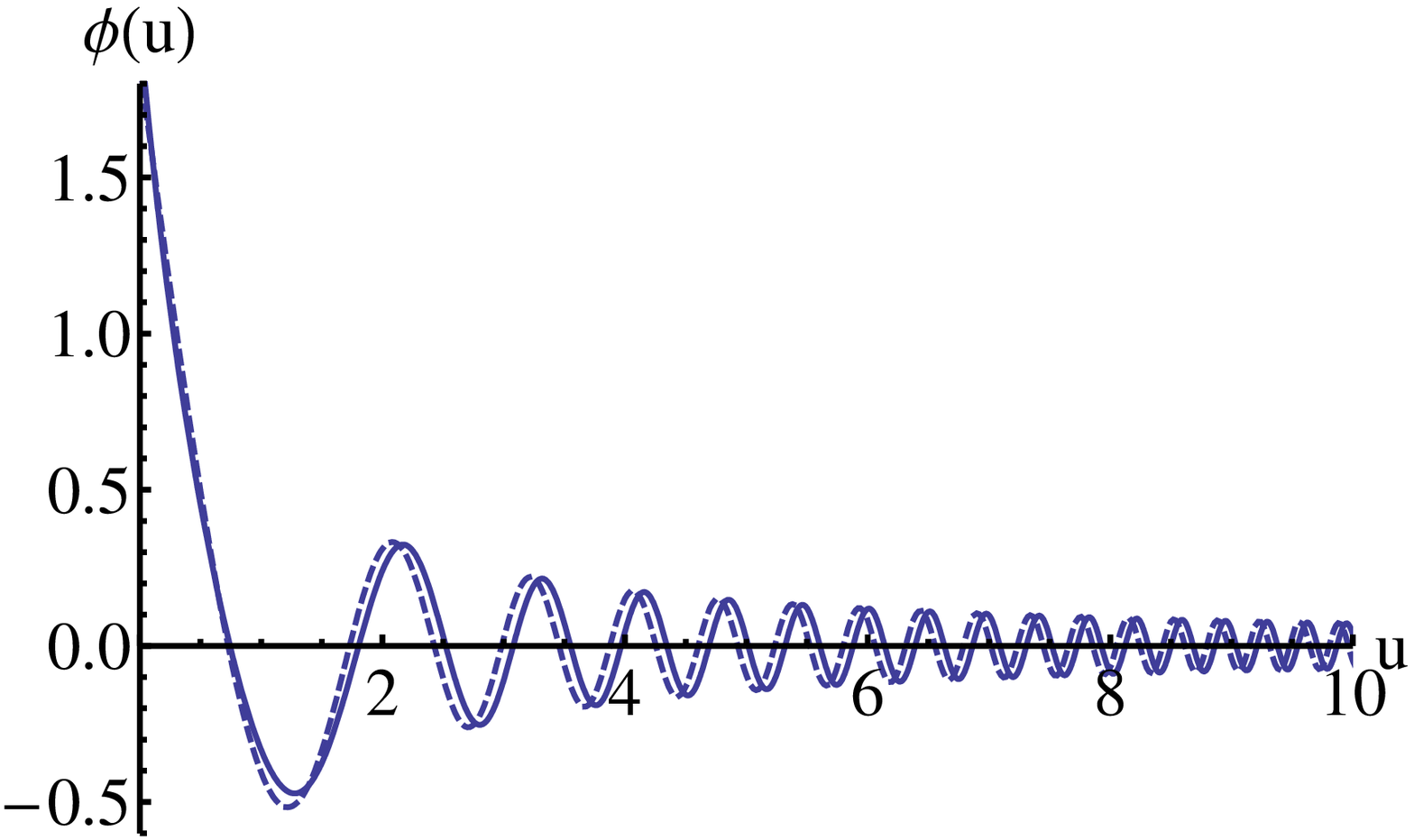}
  \caption{$\phi_{A}(u)$, for $a=0.6$, $b=2.8$ (solid line), with $\phi_{max}(u)$ for comparison (dashed line).}
\label{phi2.1}
  \end{minipage}
  \hspace{.5cm}
  \begin{minipage}[t]{0.47\linewidth}
    \centering
    \includegraphics[width=\linewidth]{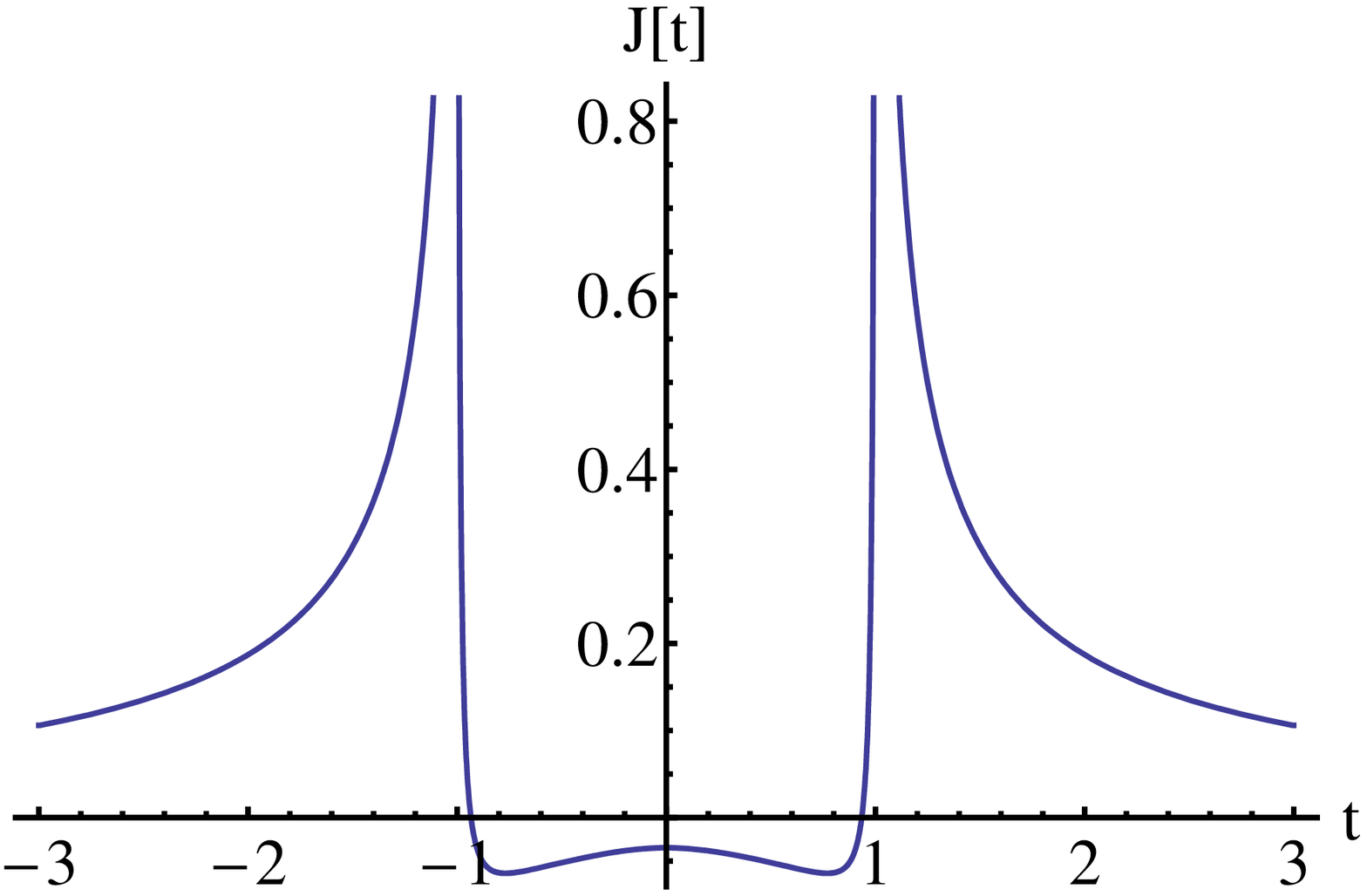}
\caption{Current, $J(t)$ for $a=0.6$ and $b=2.8$.}
\label{J2}
  \end{minipage}
\end{figure}
We plot $\phi_{A}$ in Fig.(\ref{phi2.1}) for the values of $a$ and $b$ which maximize backflow. We see good agreement with the numerical result.
The maximum amount of negative flux we can generate is
\beq
F=-0.02757,
\eeq
which occurs for the parameters $a=0.6$, $b=2.8$. This corresponds to about 70 percent of
$c_{bm}$. We plot the current $J(t)$ for these parameters in Fig.(\ref{J2}). The current is close to the numerical result, Fig.(\ref{figjexact}) away from $t=\pm1$ but lacks the correct behavior as $t\to\pm1$. This is presumably related to the fact that the asymptotic behaviour of $\phi_{A}$ does not match that of $\phi_{max}$. It would be interesting to know if there exists a better analytic approximation to the maximum backflow state that does have the correct asymptotic behavior and still gives rise to an analytically computable current. This will be explored elsewhere \cite{future}.


\section{The Classical Limit of Backflow}

Some insights into the properties of backflow may be found by looking at its classical
limit. Strictly one should do this via the usual framework of open system dynamics, and a detailed discussion of the current for an open system was given in Ref.\cite{Ye1}. In this work the resulting positivity of the current, after finite time, is clearly seen.


However it is nevertheless interesting to look at the naive classical limit of backflow characterized by the limit $\hbar \rightarrow 0$. As seen from Eq.(\ref{evalue2}) the eigenvalues of the flux operator are independent of $\hbar $. Thus although the existence of negative eigenvalues (negative flux) is clearly a quantum phenomenon,  backflow does not go away in the naive classical limit,
$\hbar \rightarrow 0$, as one might have expected. 
%
%
%
The origin of the independence of the eigenvalues of the flux operator on $\hbar$ is the fact that there is no way to construct a dimensionless number from the parameters in the problem, $\hbar$, $m$ and $T$. What is missing is a length scale. 

In any realistic experimental set up, measurements  have  finite resolution. It is therefore often more appropriate to model the measurement process using POVMs rather than exact projection operators. With this in mind, instead of defining
the flux operator in terms of exact projection operators $P = \theta (\hat x)$, instead let us define it in terms of a quasiprojector $Q$. This seems reasonable since, as discussed earlier, backflow can be measured by measuring whether the particle is in $x>0$ at two different times and, due to the inevitable imprecision of real measurements, such measurements are best modeled by quasiprojectors.
A convenient choice of quasiprojector is
\beq
Q = \int_0^{\infty} dy \ \delta_{\sigma} ( \hat x - y )
\label{quasi}
\eeq
where $\delta_{\sigma} (\hat x - y ) $ is a smoothed out $\delta$-function,
\beq
\delta_{\sigma} (\hat x - y ) = \frac {1} { (2 \pi \sigma^2)^{1/2} }
\exp \left( - \frac { (\hat x - y)^2 } {2 \sigma^2} \right)
\eeq
This goes to the usual $\delta$-function as $\sigma \rightarrow 0 $ and then $Q \rightarrow \theta (\hat x ) $.
%
%
By replacing the exact projector in the definition of the flux with this new quasi-projector we may compute modified expressions for the current and the flux \cite{HaYeBF}. In particular the eigenvalue equation Eq.(\ref{evalue2}) becomes,
\beq
\frac {1} {\pi} \int_0^\infty dv \ \frac { \sin (u^2 - v^2)  } {(u-v)}\ e^{ -a^2 (u-v)^2} \
\phi (v)
 = \lambda \phi (u)
\label{evalue4}
\eeq
where the dimensionless number $a$ is given by $ a^2 = 2 m \sigma^2 / \hbar T$. 
This means that the eigenvalues $\lambda$ will now depend on $a$, so we write $\lambda = \lambda (a)$. Through $a$ they will therefore depend on $\hbar$ and the ``limit'' $\hbar \rightarrow 0 $ now clearly means the regime $ a \gg 1 $, that is, $ \hbar \ll 2 m \sigma^2 / T $.
Hence, in a more realistic measurement situation, the bound on the total backflow -- the most negative eigenvalue of Eq.(\ref{evalue4}) -- {\it will} depend on $\hbar$ and the limit $\hbar \rightarrow 0 $ may now be more meaningful.

\begin{figure}[htbp]
\hspace{0.25\linewidth}
  \begin{minipage}[t]{0.47\linewidth}
    \centering
    \includegraphics[width=\linewidth]{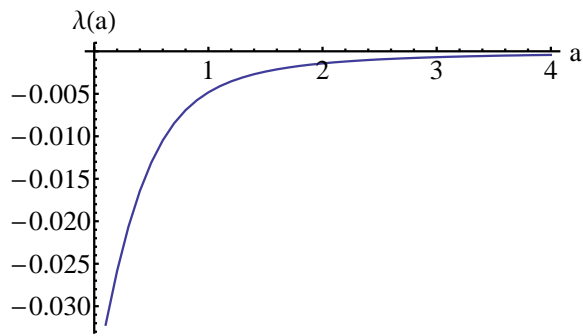}
\caption{The most negative eigenvalue of Eq.(\ref{evalue4}), $\lambda(a)$ as a function of $a$. }
\label{lambdaplot}
  \end{minipage}
  \hspace{0.25\linewidth}
\end{figure}

A reasonable conjecture is that the negative eigenvalues will increase with $a$ and also that
\beq
\lambda (a) \ge - c_{bm}
\eeq
for all $a$, so that $-c_{bm}$ emerges as a lower bound on the eigenvalues, achievable only in the limit $a \rightarrow 0 $. It seems unlikely, however, that all the negative eigenvalues will all become positive or zero, except perhaps in the limit $a \rightarrow \infty $.  We have not been able to solve Eq.(\ref{evalue4}) analytically, so instead we have obtained numerical estimates for $\lambda(a)$ for various values of $a$, and we plot the result in Fig.(\ref{lambdaplot}). The value of $\lambda(a)$ does indeed increase with $a$, tending to zero asymptotically. In fact, numerical solutions are consistent with the asymptotic form,
\beq
\lambda(a)\sim \ - \frac{1} {a^{2}} \label{lambdaas}
\eeq
for large $a$, which can be understood by examining the asymptotic form of Eq.(\ref{evalue4}) for large $a$ \cite{HaYeBF}.

In summary, for quasi-projectors, Eq.(\ref{quasi}), which are more realistic models for measurements than exact projectors, the eigenvalues, in particular the lowest eigenvalue representing the most negative flux, do depend on $\hbar$ (and $m$ and $T$). 
The lowest eigenvalue appears to go to zero for $a \rightarrow \infty$. This indicates that all negative eigenvalues go to zero (or become positive)
in the naive classical limit $\hbar \rightarrow 0 $.
By contrast the positive eigenvalues are not significantly affected \cite{HaYeBF}.

\section{Backflow and Measurement Models}

In this section we relate the above results on backflow to measurements. The aim is to begin to address the practical question of how backflow may be measured.
Many more elaborate and realistic models for the measurement of the arrival time (involving model detectors, for example) naturally lead to an arrival time probability defined with a complex potential. This is described in detail in many places \cite{HaYe1}.
These models typically yield an arrival time probability distribution which is closely related to the current and from which the current may be extracted, even when negative, thereby leading to a possible measurement of backflow.

Consider an initial wave packet starting in $x<0$ with positive momentum. We seek the arrival time probability distribution $\Pi (\tau)d \tau $ for crossing the origin between $\tau$ and $\tau + d \tau$. We consider a complex absorbing potential of step function form in $x>0$
so the Hamiltonian is $H_0 - i V_0 \theta (\hat x) $, where $H_0$ is the free Hamiltonian. We define the survival probability
$N(\tau)$ to be the norm of the state at time $\tau$ after evolution with this complex Hamiltonian.
Then
\bea
\Pi (\tau) &=& - \frac {d N} {d \tau}
\nonumber \\
&=& 2  V_0 \langle \psi | e^{ \left(  i H_0   - V_0 \theta (\hat x)   \right) \tau} \theta (  \hat x ) e^{\left( - i H_0   - V_0 \theta (\hat x)    \right) \tau } | \psi \rangle
\label{A.1}
\eea
This expression may be simplified considerably in the usual weak measurement approximation ($V_0$ small compared with the energy of the state), yielding \cite{HaYeBF},
\beq
\Pi (\tau) \approx 2  V_0 \int_{-\infty}^\tau dt \ e^{ - 2 V_0 (\tau - t) }
\ \langle \psi_t | \hat J  | \psi_t \rangle
\label{A.5}
\eeq
where $ | \psi_t \rangle = e^{-iH_0 t} | \psi \rangle $. This is the expected semiclassical result \cite{All,DEHM,HaYe1}.
Note that Eq.(\ref{A.5}) is not necessarily positive, due to the negativity of the current in certain states. However this is an artifact of the approximation and should not matter for sufficiently small $V_0$, and we will assume that Eq.(\ref{A.5}) is positive.

\begin{figure}[htbp]
\hspace{0.25\linewidth}
  \begin{minipage}[t]{0.47\linewidth}
    \centering
    \includegraphics[width=\linewidth]{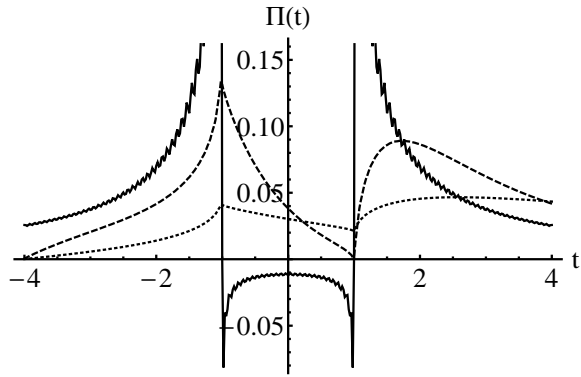}
\caption{A plot of the current (solid line) and time-smeared current Eq.(\ref{A.5}) for $V_0=0.5$ (dashed line)
   and $V_0 = 0.1 $ (dotted line).}
   \label{Fig14}
  \end{minipage}
  \hspace{0.25\linewidth}
\end{figure}

The quantity $\Pi (\tau)$ corresponds to the arrival time distribution measured by a realistic measurement so can in principle be determined experimentally. The current can then be extracted from Eq.(\ref{A.5}) by deconvolution \cite{DEHM}. 
This therefore gives a method of measuring the current and the flux, and checking for backflow.
%
%
In Fig.(\ref{Fig14}) we plot the measurement probability Eq.(\ref{A.5}) for two values of $V_0$ and also the original numerically computed current, to see how the time-smearing affects the backflow. We see that positive regions of the current are not qualitatively changed very much, in keeping with semiclassical expectations, but negative regions of the current become positive as a result of the smearing, as they must, since the measured probability is positive. Details of other possible measurements of backflow are given in Ref.\cite{HaYeBF}.


\section{Summary and Conclusions}

The purpose of this paper was to introduce various aspects of the backflow effect. After setting up the problem in Section 2, in Section 3, we showed that backflow may occur for a superposition of gaussian wavepackets. These states are important since they are experimentally realizable,  but the maximum amount of flux is very small, only about 16 percent of the maximum possible.

In Section 4, we gave an analytic expression for a state matching rather closely the numerically computed states giving maximal backflow, computed by Penz et al \cite{Penz}. The plot of the current shows reasonably good agreement with the numerical solution, except at the end points $t = \pm 1$ of the backflow region. We computed the most negative flux for this state and found it to be about 70 percent of the numerically computed maximum backflow, significantly better than any previous analytic expression for a backflow state.


In Section 5, we discussed the classical limit of backflow. The most interesting aspect of this is the issue is that the eigenvalues of the flux operator are independent of $\hbar$. This appears to mean that backflow does not go away in the naive classical limit $\hbar \rightarrow 0 $. We showed that this situation starts to appear more reasonable when the projectors used in the definition of the flux operator are replaced by quasiprojectors, which include a physical parameter characterizing the imprecision of real measurements. The eigenvalues then do depend on $\hbar$ and there is evidence that all the negative eigenvalues become zero or positive as $\hbar \rightarrow 0 $, restoring the naive classical limit.
However, there are clearly more issues to explore around this question.

In Section 6 we discussed a simple measurement model in which the current can be obtained from the measured probability by deconvolution, and from this result the negative current could in principle be obtained. The features of backflow elucidated here may be of value in designing experiments to test backflow. These and related ideas with be explored elsewhere.


\section{Acknowledgements}

We are very grateful to Gonzalo Muga for many useful conversations about the topic of this paper. JMY was supported by the John Templeton Foundation.
\vspace{10pt}

\input refs.tex


\end{document}

%% file: refs.tex
\bibliography{apssamp}